\documentclass[a4paper]{jpconf}
\usepackage{graphicx}
\usepackage[utf8]{inputenc}
\usepackage{hyperref}
\usepackage{cite}
\usepackage{iopams}
\newcommand{\sgn}{\mathrm{sgn}}
\begin{document}
\title{Beyond Bell's theorem: Admissible hidden-variable models for the spin-singlet}
\author{Antonio Di Lorenzo}
\address{Instituto de F\'{\i}sica, Universidade Federal de Uberl\^{a}ndia, 38400-902 Uberl\^{a}ndia, Minas Gerais, Brazil}
\ead{dilorenzo@infis.ufu.br}
\begin{abstract}
Assuming that quantum mechanics is obeyed exactly after averaging over hidden variables, 
and considering models that obey both the hypotheses of free will and locality, 
we establish the form of all possible hidden-variable models that reproduce the spin-singlet.  
\end{abstract}

\section{Introduction}
Bell \cite{Bell1964} established an inequality, which was later extended and reformulated \cite{Clauser1969,Clauser1974}, obeyed by a large family of theories, deterministic or statistical, based on the assumption that 
the quantum state is not the ultimate limit, but that there may be additional parameters, the hidden variables, providing a finer description. 
Experiments do not only violate the inequality, but they also reproduce the quantum mechanical predictions with high 
precision and accuracy \cite{Freedman1972,Aspect1982a,Aspect1982b,Tapster1994,Weihs1998,Tittel1998,Rowe2001,
Groblacher2007,Paterek2007,Branciard2007,Eisaman2008,Romero2010,Paternostro2010,Lee2011}. 
Furthermore, the empirical data show an increasing agreement with quantum mechanics as 
experimental techniques become more refined. 

Here, we shall extrapolate this trend, and assume that quantum mechanics is obeyed exactly, in absence of any knowledge 
about the hidden variable. 
This puts further restrictions on the admissible hidden-variable theories, as they must not only violate Bell inequality, 
but also reproduce quantum mechanics after the hidden variables are averaged out. 
Under the hypotheses that the hidden-variable models satisfy the locality condition and the free will assumption,  
it is possible to determine their shape, thus going beyond Bell's theorem.

The main result of this work was published in Ref.~\cite{DiLorenzo2012c}. 
Here, we just give a more compact account of the result, while dedicating some more space to epistemological issues. 
\section{Hypotheses at the basis of Bell inequality}
Bell inequality follows easily under the assumption that the joint probability $P_{\sigma,\tau}(\bi{a},\bi{b},\psi)$ of observing an outcome $\sigma\in\{-1,1\}$ at 
a detector characterized by $\bi{a}$ and $\tau\in\{-1,+1\}$ at a detector characterized by $\bi{b}$, for a fixed preparation $\psi$, can be 
decomposed as 
\begin{equation}\label{eq:bellfact}
P_{\sigma,\tau}(\bi{a},\bi{b},\psi) = \int \rmd\mu(\lambda|\psi) M_{\sigma}(\bi{a},\lambda,\psi) M_{\tau}(\bi{b},\lambda,\psi) ,
\end{equation}
with $\mu$ a positive distribution normalized to one, and $M$ marginal probability conditioned on $\lambda$. 
In the following, we shall consider implicit for brevity the dependence on $\psi$, as the preparation does not change.   
We remark that applying Bayes' rule, the probability generally factors as 
\begin{equation}\label{eq:bayesfact}
P_{\sigma,\tau}(\bi{a},\bi{b}) = \int \rmd\mu(\lambda|\bi{a},\bi{b}) M_{\sigma}(\bi{a},\bi{b},\lambda) Q_{\tau}(\sigma,\bi{a},\bi{b},\lambda) ,
\end{equation}
with $Q$ conditional probability for given $\sigma$. 
In order to comply with Eq.~\eref{eq:bellfact}, the models considered by Bell must satisfy three hypotheses: 
$\mu(\lambda|\bi{a},\bi{b})=\mu(\lambda)$, often justified on the basis of ``free will",  but which we shall address with the technical term Uncorrelated Choice;   
$M_{\sigma}(\bi{a},\bi{b},\lambda)=M_{\sigma}(\bi{a},\lambda)$ and 
$M_{\tau}(\bi{a},\bi{b},\lambda)=M_{\tau}(\bi{b},\lambda)$, called 
``Setting-Independence'', sometimes identified with the no-signaling principle, mistakenly;  
$Q_{\tau}(\sigma,\bi{a},\bi{b},\lambda)=
M_{\tau}(\bi{a},\bi{b},\lambda)$, called in the literature Outcome-Independence, which we shall call 
``Reducibility of Correlations".

Let us discuss these assumptions. 
\subsection{Uncorrelated Choice (UC)}
This assumption means that the choice of measurement and the distribution of the hidden variables $\lambda$  are uncorrelated, 
so that there is no causal connection between the two. 
Let us assume momentarily that UC does not hold. 
If one thinks of $\lambda$ as a set of parameters attached to 
the physical system, and assumes that the choice of measurement $\Sigma=\{\bi{a},\bi{b}\}$ is the cause and the change in the distribution 
$\mu(\lambda|\Sigma)$ is the effect, 
then a violation of Uncorrelated Choice would result in non-locality.
 However, if one inverts the cause-effect relationship, then a violation of Uncorrelated Choice 
corresponds to a limitation of free will \cite{Brans1988,Hall2010,DiLorenzo2012b,DiLorenzo2013a}. 
As an example, think of an astronaut on a distant planet, who is video-communicating with an engineer at NASA. 
You notice that, after subtracting the time delay due to the finite speed of communication,  
when the astronaut raises her left arm, the engineer does as well, and so when she scratches her head, etc. 
What could you conclude by observing such correlations? Perhaps the astronaut has access to an image of the engineer through some superluminal channel, and is just mimicking the gestures she can see in advance, or vice versa; 
perhaps the astronaut and the engineer have agreed to make those gestures at those times, in the first interplanetary flashmob ever; or perhaps they have no free will, and 
the same forces that determine his movements are determining hers.
Thus, as pointed out e.g. in Refs.~\cite{DiLorenzo2012b,DiLorenzo2013a}, there is no way one can decide whether the correlations are due to conspiracy, non-locality, or slave will, unless a physical theory is formulated that describes how 
the decisions to do a movement (i.e., discontinuing the metaphor, to choose a measurement) are taken.

Marketing experts and statisticians can confirm that people who are exposed to the same advertisements or who 
share the same social conditions tend to buy the same products or to vote for the same candidate. 
Thus, we could accept a limitation of free will. However, with respect to the hypothesis of Uncorrelated Choice, 
we make the following consideration: 
the experimenters deciding the settings $\Sigma$ can delegate the choice to third parties, as for instance 
the fluctuations in the voltage of an electric socket, the variations in the albedo of a distant star, etc.  
As there is empirical evidence \cite{Scheidl2010} that Bell inequality is violated in this case as well, the only possible 
conclusion, if one insists that UC is violated, is that there is a leibnitzian pre-established harmony, where the hidden variables determine not only 
the choice of $\Sigma$, but also the choice of an apparent random process among a multitude of possible ones. 
Furthermore, the outcome of this apparently random process is determined as well by $\lambda$. 
Thus, while we cannot rule out a violation of free will as an explanation, we can say, based on partial 
experimental evidence, that if this is the case then we should admit a cosmic choreography. 
As far as a theory of the intentions and preferences of the choreographer is not put forward, 
the assumption is untestable, as it can explain anything and predict nothing, 
and thus we should store it away, together with solipsism and intelligent design. 
In conclusion, we do assume that Uncorrelated Choice holds. 

\subsection{Setting-Independence (SI)}
This hypothesis requires that the marginal probability of observing an event 
$\sigma$ in a space-time region $A$, 
for a given $\lambda$, does not depend on the settings $\Sigma'$ at space-time regions outside of the 
lightcone (past and future) of $A$. 
Often, SI is considered as synonymous of no-signaling, a form of locality,\footnote{Here, we consider locality and no-signaling as synonymous, all other definitions of locality 
being unsatisfactory, as discussed below in the main text.} in the sense that if SI holds, then it is impossible to send 
superluminal signals from a region $B$ to a region $A$. Actually SI is weaker than no-signaling, as SI may hold but signaling still be possible. 
Let us see how: suppose that the hidden variables $\lambda$ can be divided in $\lambda_A$, a set of parameters localized in region $A$, $\lambda_B$, localized in region $B$, 
and possibly $\lambda_G$, a set of global parameters that cannot be associated to any particular region (the wave function of entangled particles is an example of 
a global parameter). The assumption SI allows the marginal probability at $A$ to depend on $\lambda_B$, 
$M_{\sigma}(\bi{a},\bi{b},\lambda_A,\lambda_B,\lambda_G)=M_{\sigma}(\bi{a},\lambda_A,\lambda_B,\lambda_G)$. 
Thus, if $\lambda_B$ changes, so does the marginal probability at $A$. By receiving a large number of particles in parallel and estimating 
the marginal probability by the frequency of observing $\sigma$, an observer at $A$ can thus receive an instantaneous message from an observer at $B$, provided 
the latter has a way to influence $\lambda_B$. 
To me, the existence of this non-locality is dubious, and therefore I shall assume that SI holds, and furthermore that the marginal probability does not depend on $\lambda_B$. 

A special case of SI is the compliance with Malus's law, 
postulated by Leggett in the derivation of a new inequality \cite{Leggett2003}. It requires that the hidden-variables consist in two unit vectors $\bi{u}$ and $\bi{v}$,  
such that the marginal probabilities are 
$M_\sigma(\bi{u},\bi{a},\bi{b})=\left(1+\sigma\bi{a}\cdot\bi{u}\right)/2$ and 
$M_\tau(\bi{v},\bi{a},\bi{b})=\left(1+\sigma\bi{b}\cdot\bi{v}\right)/2$. 
There is a result \cite{Colbeck2008,Branciard2008,DiLorenzo2012c}, which we may call the \emph{trivial-marginals theorem}, 
establishing that all theories that obey UC and SI, and that also reproduce the quantum mechanical predictions for a spin singlet must have 
a marginal probability $M_\sigma(\lambda,\bi{a},\bi{b})=1/2$. 
Thus, the compliance with Malus's law is excluded, and in the following we shall use this theorem, which also guarantees the locality 
of the models. 
\subsection{Reducibility of Correlations (RC)} This assumption, called outcome-independence by Shimony \cite{Shimony1990},  states that 
$Q_\tau(\sigma,\bi{a},\bi{b},\lambda)=M_\tau(\bi{a},\bi{b},\lambda)$, 
i.e. the conditional probability of obtaining an outcome $\tau$, for fixed $\lambda$ and given that the outcome of the remote measurement along $\bi{a}$ is $\sigma$, 
does not depend on the latter outcome. 
Hence, after applying Bayes' rule, the joint probability factorizes symmetrically 
$P_{\sigma,\tau}(\bi{a},\bi{b},\lambda)=
M_\sigma(\bi{a},\bi{b},\lambda) M_\tau(\bi{a},\bi{b},\lambda)$.
Hence, for fixed parameters $\lambda$, there would be no correlations, and these would appear only 
after averaging over $\lambda$. Thus the assumption of Reducibility of Correlations amounts to state 
that the quantum correlations are a result of the ignorance of more fundamental parameters.

The hypothesis RC is implied by determinism, in the sense that, if by knowing $\lambda$ one may predict with certainty 
whether the outcome $\tau$ will occur or not, then the knowledge of the outcome $\sigma$ is superfluous. 
While the vice versa is not true, however, a theorem established by M. Hall \cite{Hall2011} shows that any model obeying RC admits 
a natural deterministic completion, i.e. a description in terms of additional variables $\mu$ such that 
$M_\sigma(\bi{a},\bi{b},\lambda,\mu)\in\{0,1\}$. Thus, RC generalizes the concept of determinism. 

In order to verify whether Reducibility of Correlations holds, an observer in the region $A$ should ask the observer at $B$ about the $\tau$ he measured and the $\bi{b}$ he used, 
then evaluate the conditional probability $Q_\sigma(\tau,\bi{a},\bi{b},\lambda)$, and finally check whether this quantity changes with $\tau$, with the remaining parameters fixed. In order to do so, $A$ must have access to the remote information $\bi{b},\tau$, which $B$ can send only at  a speed not exceeding the speed of light.
Hence, violating Reducibility of Correlations 
does not imply action-at-a-distance, nor the possibility of instantaneous communication. 
Yet, the locution ``quantum nonlocality'' is widely used. The reason is that originally Bell \cite{Bell1964} did assume determinism, 
thus justifying RC, and locality. Later generalizations of Bell inequality \cite{Bell1971,Clauser1974} 
dropped the hypothesis of determinism, but kept assuming RC, in the mistaken belief that this was a consequence of locality, not of determinism.
Jarrett \cite{Jarrett1984}, and subsequently Shimony \cite{Shimony1990}, pointed out that RC does not follow from the notion of locality, as commonly meant. 
This resulted in some new terminology being introduced to refer to both the hypothesis SI and RC: strong locality \cite{Jarrett1984}, Bell locality, local causality \cite{Bell1990}.  
After a while, this distinction fell, and unfortunately many studies of Bell inequality refer to the two hypotheses SI and RC (and sometimes UC) as ``locality'', while 
the reader of the present manuscript should be aware that this is just a misnomer: there is no factual evidence of an intrinsic non-locality in Nature. 
Furthermore, the presumed non-locality of Nature (or of quantum mechanics) would be 
of a very peculiar kind, as it would manifest in the kinematics. By this, we mean that, since ordinary quantum mechanics is, as classical mechanics, a non-relativistic theory, 
it can accept non-local forces (think, e.g., to Newtonian gravitation or Coulomb's law). However, the ``non-locality'' of entangled states manifests when there is no known 
interaction between the two subsystems, and thus it would be a feature of the very kinematics of the theory, i.e. we would have ``non-locality'' without interaction, a 
patent absurdity.

In the following, we shall assume that the hypothesis RC does not hold, as this is the only assumption not fully justified by physical considerations.  
\section{Models for the spin singlet}
As one or more of the three hypotheses discussed in the previous section are dropped, it is possible to violate Bell inequality. 
However, this is a necessary but not sufficient condition in order to reproduce experimental data, as these agree well with the predictions of quantum mechanics.  
In the Appendix, we review some significant models that reproduce the probability for the spin singlet after averaging out the hidden variables. 
None of these models relies solely on the violation of Reducibility of Correlations.

\section{Main theorem}
As our goal is to find models that reproduce the quantum mechanical predictions, $P_{\sigma,\tau}(\bi{a},\bi{b})=(1-\sigma\tau \bi{a}\cdot\bi{b})/4$,  
and that obey both UC and SI, 
we can apply the trivial-marginals theorem, so that the joint probability can be written as 
\begin{equation}
P_{\sigma,\tau}(\bi{a},\bi{b}) = \int \rmd\mu(\lambda) \frac{1}{4} \left[1+\sigma\tau D(\lambda,\bi{a},\bi{b})
\right].
\end{equation}
The question that we ask is: what functions $D$ are admissible (i.e. give a probability between 0 and 1) so that the quantum mechanical predictions are 
reproduced after averaging out $\lambda$? 
The main idea is of using the tools of Calculus to approach this problem, while former approaches rely on algebra. 
Without loss of generality we can write 
\begin{equation}
D(\lambda,\bi{a},\bi{b}) = -\bi{a}\cdot\bi{b}+C(\lambda,\bi{a},\bi{b})
\end{equation}
Clearly, $\int \rmd\mu(\lambda) C(\lambda,\bi{a},\bi{b})=0$ identically in $\bi{a},\bi{b}$, so that $C$ must take both positive and negative values. 

Furthermore, let us take $\sigma=\tau$ and $\bi{a}=\bi{a}$. 
Then the probability, conditioned on $\lambda$, reduces to $P_{\sigma,\sigma}(\lambda,\bi{a},\bi{a})=C(\lambda,\bi{a},\bi{a})$, 
and this implies that $C(\lambda,\bi{a},\bi{a})=0, \forall \bi{a},\lambda$. Analogously, you can prove that 
$C(\lambda,\bi{a},-\bi{a})=0, \forall \bi{a},\lambda$. 
Thus, perfect correlations and anti-correlations limit the admissible $C$ to the form
\begin{equation}
C(\lambda,\bi{a},\bi{b})=
\left[1+\bi{a}\cdot\bi{b}\right]^{s_+}\left[1-\bi{a}\cdot\bi{b}\right]^{s_-} 
G(\lambda,\bi{a},\bi{b}), 
\end{equation}
with $s_+>0,s_->0$ and $0<|G(\lambda,\bi{a},\pm\bi{a})|<\infty$. 
This is but a Frobenius expansion, with the indices $s_\pm$ 
determining how fast the function vanishes close to the perfect (anti-)correlations configuration $\bi{a}\cdot\bi{b}=\mp 1$.  
Compatibility with quantum mechanics requires 
\begin{equation}
\int \rmd\mu(\lambda) G(\lambda,\bi{a},\bi{b}) =0, \forall \bi{a}, \bi{b}.
\end{equation} 

Now, let us consider a configuration close to the perfect anticorrelation $\bi{a}\cdot\bi{b}=1-\bi{\varepsilon}$, with $\varepsilon\ll 1$. 
Then 
\begin{equation}\label{eq:exp}
P_{\sigma,\sigma}(\bi{a},\bi{b},\lambda)\simeq\frac{\varepsilon}{4}\left[
1- 2^{s_+}\varepsilon^{s_{-}-1} G(\lambda,\bi{a},\bi{a})\right] , 
\end{equation}
The positivity of the probability implies that $s_-\ge 1$. The inequality $s_+\ge 1$ is proved analogously.

\section{Conclusions}
In conclusion, we have established the form of all the hidden variable models able to reproduce the quantum mechanics of a spin-singlet by satisfying both 
the assumptions of \lq\lq{}free will\rq\rq{} and no-signaling, which correspond, respectively, to 
Uncorrelated Choice (Measurement-Independence) and Setting-Independence. 
By contrast, we have assumed the violation of Reducibility of Correlations, as this can never result in superluminal signaling, since it consists in   
the dependence of a conditional probability on a remote outcome, and as such it requires the communication 
of said outcome through means that are necessarily subluminal. 
Rather, the violation of Reducibility of Correlations implies that the quantum correlations cannot be 
attributed to the ignorance of the hidden parameters.

\ack
This work was performed as part of the Brazilian Instituto
Nacional de Ci\^{e}ncia e Tecnologia para a Informa\c{c}\~{a}o Qu\^{a}ntica 
(INCT–IQ) and it was supported by Funda\c{c}\~{a}o de Amparo \`{a} Pesquisa do 
Estado de Minas Gerais through Processes No. APQ-02804-10 and PEE-00421-12.

\appendix 
\section*{Appendix: Review of models reproducing the quantum-mechanical correlations}
\setcounter{section}{1}
In the literature there is a handful of hidden-variable models 
able to reproduce quantum mechanics for a spin-singlet. These models must violate 
at least one of Bell's hypotheses. 
However, as it is not always clear which hypotheses are dropped, 
we shall briefly review the existing (to the best of our knowledge) models. 
We recall that a hidden-variable theory predicts the decomposition of the quantum probability 
as 
 \begin{equation}\label{eq:deco}
\int \rmd\mu(\lambda|\bi{a},\bi{b}) P_{\sigma,\tau}(\lambda,\bi{a},\bi{b})= 
\frac{1}{4}\left[1-\sigma\tau \bi{a}\cdot\bi{b}\right].
\end{equation}
Inspection of $\mu(\lambda|\bi{a},\bi{b})$, depending whether it depends 
on $\bi{a},\bi{b}$ or not, allows to establish if Uncorrelated Choice holds. 
After extracting the marginal probability $M_{\sigma}(\lambda,\bi{a},\bi{b})$ 
from $P_{\sigma,\tau}(\lambda,\bi{a},\bi{b})$, one can check whether 
Setting-Independence (and possibly Malus's law) is satisfied. 
Finally, obtaining the conditional probability 
$Q_{\sigma}(\lambda,\bi{a},\bi{b},\tau)$, the hypothesis of Reducibility of Correlations 
can be checked. 

\subsection{Brans model.}
Brans \cite{Brans1988} proposed the following model using four discrete binary hidden-variables 
and two unit-vectors: 
\begin{eqnarray}
\mu(j,k,\alpha,\beta,\bi{u},\bi{v}|{\bi{a},\bi{b}})=
\frac{jk}{2} \delta(\bi{u}-\bi{a})\delta(\bi{v}-\bi{b})
\langle j_{\hat{\bi{z}}}|\alpha_{\bi{u}}\rangle
\langle \alpha_{\bi{u}}|k_{\hat{\bi{z}}}\rangle 
\langle -j_{\hat{\bi{z}}}|\beta_{\bi{v}}\rangle
\langle \beta_{\bi{v}}|-k_{\hat{\bi{z}}}\rangle ,
\label{eq:Bransmu}
\\
P_{\sigma,\tau}(j,k,\alpha,\beta,\bi{u},\bi{v},\bi{a},\bi{b})= 
\frac{1}{4}(1+\sigma \alpha) (1+\tau\beta).
\label{eq:Bransp}
\end{eqnarray}
The symbol $|\alpha_{\bi{u}}\rangle$ denotes the eigenstate $\alpha=\pm 1$ of the 
spin component $\bi{u}\cdot\boldsymbol{\sigma}$. 
Thus, Brans's model violates only the Uncorrelated Choice hypothesis. 
Using the standard definition of spin eigenstates that can be found in Sakurai's textbook \cite{Sakuraitextbook}, 
after some algebra one gets 
\begin{eqnarray}
\fl \mu(j,k,\alpha,\beta,\bi{u},\bi{v}|\bi{a},\bi{b})=
\frac{1}{2}\delta(\bi{u}-\bi{a})\delta(\bi{v}-\bi{b})
\nonumber
\\
\nonumber
\times 
\biggl\{
\delta_{j,k} 
\left[\delta_{j,\alpha}\cos^2{(\theta_\bi{u}/2)}+
\delta_{j,-\alpha} \sin^2{(\theta_\bi{u}/2)}\right]
\left[\delta_{j,-\beta}\cos^2{(\theta_\bi{v}/2)}+
\delta_{j,\beta} \sin^2{(\theta_\bi{v}/2)}\right]
\\
\label{eq:negpart}
-\delta_{j,-k} \frac{\alpha\beta}{4}  \sin{\theta_\bi{u}}\sin{\theta_\bi{v}}
e^{-i j(\phi_\bi{u}-\phi_\bi{v})}
\biggr\},
\end{eqnarray}
with $0\le \theta_\bi{u},\theta_\bi{v}\le \pi$ and $\phi_\bi{u},\phi_\bi{v}$
the azimuthal and polar angles of the unit vectors 
in the arbitrary coordinate system fixing the phase of the eigenstates. 
Clearly, Eq.~\eref{eq:negpart} is non-real, and gives a negative probability for 
$\alpha=\beta$, $j=-k$, $\bi{u}=\bi{v}$. 
We notice however that $j,k$ do not appear in the conditional probability Eq.~\eref{eq:Bransp}, 
thus, after summing over them, we have 
\begin{eqnarray}
\nonumber
\mu(\alpha,\beta,\bi{u},\bi{v}|\bi{a},\bi{b})=&&\!\!\!\!\!\!\!\!
\delta(\bi{u}-\bi{a})\delta(\bi{v}-\bi{b})
\frac{1-\alpha\beta\left[\cos{\theta_\bi{u}} \cos{\theta_\bi{v}}
-  \sin{\theta_\bi{u}}\sin{\theta_\bi{v}} \cos{(\phi_\bi{u}-\phi_\bi{v})}\right]}{4}\\
\label{eq:Branscorr}
=&&\!\!\!\!\!\!\!\!
 \frac{1}{4} \delta(\bi{u}-\bi{a})\delta(\bi{v}-\bi{b}) 
\left[1-\alpha\beta \bi{u}\cdot\bi{v}\right].
\end{eqnarray}
This is the correct, positive-definite distribution to replace Eq.~\eref{eq:Bransmu}. 
In conclusion, Brans's model, properly formulated in terms of a real and positive distribution, 
turns out to be trivial: 
there are four hidden variables $\alpha,\beta,\bi{u},\bi{v}$, distributed 
according to Eq.~\eref{eq:Branscorr}, so that the conditional distribution of $\alpha,\beta$ 
for given $\bi{u},\bi{v}$ mimics by construction the quantum mechanical 
distribution, $\mu(\alpha,\beta,\bi{u},\bi{v})=(1-\alpha\beta\bi{u}\cdot\bi{v})/4$. 
The variables $\alpha,\beta$ are fixed by Eq.~\eref{eq:Bransp} to coincide with the observed 
values of $\sigma,\tau$, while $\bi{u},\bi{v}$, in virtue of Eq.~\eref{eq:Branscorr} 
coincide with $\bi{a},\bi{b}$. 
\subsection{Toner and Bacon model.}\label{app:tb}
In Ref.~\cite{Toner2003}, the hidden variables are a pair of unit vectors $(\bi{u},\bi{v})$, each uniformly distributed over the unit-sphere.  
The probabilities characterizing the model are
\begin{eqnarray}
\label{eq:tbmu}
\mu(\bi{u},\bi{v}|\bi{a},\bi{b})=\frac{1}{(4\pi)^2},\\
\label{eq:tbp}
P_{\sigma,\tau}(\bi{u},\bi{v}, \bi{a},\bi{b})=
\frac{1}{4}\left[1+\sigma\mathrm{sgn}(\bi{u}\cdot\bi{a})\right]
\left\{1+\tau\mathrm{sgn}\left[\bi{u}\cdot\bi{b}+\mathrm{sgn}(\bi{u}\cdot\bi{a})\mathrm{sgn}(\bi{v}\cdot\bi{a})\bi{v}\cdot\bi{b}\right]\right\}.
\end{eqnarray}
The model is clearly deterministic (we wrote $\delta_{a,b}=(1+ab)/2$ for $a,b=\pm1$ to avoid clumsy indices) 
and asymmetric, since it can be realized by $A$ making a measurement which decides the output $\sigma$ depending on the sign of the projection of 
$\bi{u}$ over $\bi{a}$, and by subsequently $A$ sending a bit 
\begin{equation}\label{eq:TBX}
c=\mathrm{sgn}(\bi{u}\cdot\bi{a})\mathrm{sgn}(\bi{v}\cdot\bi{a})
\end{equation}
to $B$, where the output $\tau$ is determined by a suitable combination of $\bi{b},\bi{u},\bi{v}$ and $c$. 
The asymmetry yields one marginal probability obeying Setting-Independence, and the other not obeying:
\begin{eqnarray}
\label{eq:tbmarga}
M^{(A)}_{\sigma}(\bi{u},\bi{v}, \bi{a},\bi{b})=
\frac{1}{2}\left[1+\sigma\mathrm{sgn}(\bi{u}\cdot\bi{a})\right] ,\\
\label{eq:tbmargb}
M^{(B)}_{\tau}(\bi{u},\bi{v}, \bi{a},\bi{b})=
\frac{1}{2}\bigl\{1+\tau
\mathrm{sgn}\left[\bi{u}\cdot\bi{b}+\mathrm{sgn}(\bi{u}\cdot\bi{a})\mathrm{sgn}(\bi{v}\cdot\bi{a})\bi{v}\cdot\bi{b}\right]\bigr\}.
\end{eqnarray}
 Equation \eref{eq:tbmu} guarantees that Uncorrelated Choice holds, while 
the factorizability of Eq.~\eref{eq:tbp} implies the validity of Reducibility of Correlations.
\subsection{Cerf et al. model.}\label{app:cerf}
Cerf and collaborators \cite{Cerf2005} proposed a model relying on 
the Popescu-Rohrlich box \cite{Popescu1994}. 
The latter is a hypothetical machine taking two binary inputs $x=\pm 1$ and $y=\pm 1$ 
from location $A$ and $B$, resp., and 
giving correlated outputs at locations $A$ and $B$, according to 
\begin{equation}
\label{eq:prbox}
4-2|o_A+o_B|=(1-x)(1-y).
\end{equation}
 The model of Cerf \emph{et al.} assumes  
as hidden variables $x,y$ and two unit vectors $\bi{u},\bi{v}$, 
both vectors being accessible through shared randomness\footnote{We notice, however, 
that classical resources cannot provide true shared randomness, but only a simulation of it 
by means of two identical random generators synchronized and shared between $A$ and $B$. 
The Kolmogorov complexity of the sequence of outputs would result less than 
the one of true random events  to an observer intelligent enough to guess the algorithm.} 
by $A$ and $B$. The model relies on an anthropomorphic description, but can be 
stated in a neutral form: the observer at $A$ checks $\bi{u},\bi{v}$, 
arbitrarily chooses $\bi{a}$, and then inputs 
\begin{equation}
\label{eq:Cerfx}
x=\sgn{(\bi{u}\cdot\bi{a})}\ \sgn{(\bi{v}\cdot\bi{a})},
\end{equation} 
where $\sgn{(t)}=1$ for non-negative $t$, $\sgn{(t)}=-1$ for negative $t$; 
analogously the observer at $B$, after arbitrarily choosing $\bi{b}$, 
will input 
\begin{equation}
\label{eq:Cerfy}
y=\sgn{(\bi{n}_+\!\cdot\!\bi{b})}\ \sgn{(\bi{n}_-\!\cdot\!\bi{b})},
\end{equation} 
with $\bi{n}_\pm=\bi{u}\pm\bi{v}$. 
The Popescu-Rohrlich box will give then outputs according to 
Eq.~\eref{eq:prbox}, which are transformed to the simulated outcomes 
$\sigma=-o_A\ \sgn{(\bi{u}\cdot\bi{a})}$, 
$\tau=o_B\ \sgn{(\bi{n}_+\cdot\bi{b})}$. 
Assuming a uniform distribution of $\bi{u},\bi{v}$, 
the quantum mechanical correlator is recovered.  

Restating the formulation in a neutral form, we have that Cerf \emph{et al.} 
model is 
\begin{eqnarray}
\label{eq:Cerfmu}
\mu(x,y,\bi{u},\bi{v}|\bi{a},\bi{b})=
\frac{1}{(4\pi)^2}\, 
\delta\!\left[x\!-\!\sgn{(\bi{u}\!\cdot\!\bi{a})}\ \sgn{(\bi{v}\!\cdot\!\bi{a})}\right] 
\delta\left[y\!-\!\sgn{(\bi{n}_+\!\cdot\!\bi{b})}\ \sgn{(\bi{n}_-\!\cdot\!\bi{b})}\right],\\
P_{\sigma,\tau}(x,y,\bi{u},\bi{v},\bi{a},\bi{b})=
\frac{1}{4}\biggl\{1-\sigma\tau\frac{1+x+y-xy}{2}
 \sgn{(\bi{u}\cdot\bi{a})}\sgn{(\bi{n}_+\cdot\bi{b})}\biggr\}	
\label{eq:Cerfp}
.
\end{eqnarray}
Thus the model considered violates both Uncorrelated Choice and 
Reducibility of Correlations. 
The presentation of the model, however, might obscure its characteristics, 
since it is stated that the choices $\bi{a},\bi{b}$ are made freely. 
We remark that 
Eqs.~\eref{eq:Cerfx} and \eref{eq:Cerfy} constrain the values of $x,y$. 
Mathematics and probability theory are time-symmetric and acausal, 
the first because of the symmetry of the equality sign, the second because it provides correlations. 
Thus, the interpretation of what is the cause and what the effect is purely metaphysical. 
One may interpret Eqs.~\eref{eq:Cerfx} and \eref{eq:Cerfy}, or Eq.~\eref{eq:Cerfmu}, 
as limiting the possible choices 
of $\bi{a},\bi{b}$ for given $x,y$. 
In this case the model can be realized through classical resources, since the inputs 
$x,y$ are given, and the observers are instructed, or programmed, to choose only 
the $\bi{a},\bi{b}$ satisfying Eqs.~\eref{eq:Cerfx} and \eref{eq:Cerfy}. 

The model can be reduced to one violating only the Reducibility of Correlations 
hypothesis after integrating out $x$ and $y$ (thus the PR box is just a mathematical 
artifact to justify the procedure): 
\begin{eqnarray}
\label{eq:Cerfmu2app}
\mu(\bi{u},\bi{v}|\bi{a},\bi{b})=
\frac{1}{(4\pi)^2},\\ 
P_{\sigma,\tau}(\bi{u},\bi{v},\bi{a},\bi{b})=
\frac{1}{4}\biggl\{\!1\!-\!\sigma\tau\sgn{(\bi{u}\cdot\bi{a})}\sgn{(\bi{n}_+\!\cdot\!\bi{b})}
 \frac{1+x_{\bi{a},\bi{u},\bi{v}}+y_{\bi{b},\bi{u},\bi{v}}
-x_{\bi{a},\bi{u},\bi{v}}y_{\bi{a},\bi{u},\bi{v}}}{2} \biggr\},
\label{eq:Cerfp2app}
\end{eqnarray}
where $x_{\bi{a},\bi{u},\bi{v}}$ and $y_{\bi{b},\bi{u},\bi{v}}$
are the right-hand-sides of Eqs.~\eref{eq:Cerfx} and \eref{eq:Cerfy}. 
This corresponds to a $C$ function (see main text): 
\begin{equation}
C(\bi{u},\bi{v},\bi{a},\bi{b})= \bi{a}\cdot\bi{b}-\sgn{(\bi{u}\cdot\bi{a})}\sgn{(\bi{n}_+\cdot\bi{b})}
\frac{x_{\bi{a},\bi{u},\bi{v}}+y_{\bi{b},\bi{u},\bi{v}}-x_{\bi{a},\bi{u},\bi{v}}y_{\bi{b},\bi{u},\bi{v}}-1}{2} .
\end{equation}
We leave to the reader the tedious verification that 
$C(\bi{u},\bi{v},\bi{a},\pm\bi{a})=0$. 
Furthermore, because of the piecewise constancy of the $\sgn$ function, 
$C(\bi{u},\bi{v},\bi{a},\bi{b})\displaystyle{\mathop{=}_{\bi{b}\to\pm\bi{a}}} \bi{a}\cdot\bi{b}\mp 1$, 
identically for almost all values of $\bi{a}$ 
(the exception are the values $\bi{a}=\bi{u},\bi{v},\bi{u+v},\bi{u-v}$)
so that we have, with reference to our main theorem,  
$s_+=s_-=1$. 
\subsection{Gr\"{o}blacher et al. model.}
In Ref. [\citeonline{Groblacher2007b}] 
the following model, relying on two unit vectors as hidden 
variables,\footnote{The original formulation is more contrived, 
since it introduces additional variables which, in the end, are integrated over.} 
was proposed\begin{eqnarray}
\label{eq:Grobmu}
\mu(\bi{u},\bi{v}|\bi{a},\bi{b})=0,\ \mbox{for}\  
|\bi{u}\cdot \bi{a}\pm \bi{v}\cdot \bi{b}|>1\mp \bi{a}\cdot \bi{b} ,\\
\label{eq:Grobp}
P_{\sigma,\tau}(\bi{u},\bi{v},\bi{a},\bi{b})=
\frac{1}{4}\left[1\!+\!\sigma \bi{u}\cdot\bi{a}
\!+\!\tau \bi{v}\cdot\bi{b}
-\sigma\tau \bi{a}\cdot \bi{b}\right]
.
\end{eqnarray}
The distribution $\mu$, besides satisfying \eref{eq:Grobmu}, should also comply with 
$\int d\bi{u}d\bi{v} \mu(\bi{u},\bi{v}|\bi{a},\bi{b}) \bi{u}=0$ and
$\int d\bi{u}d\bi{v} \mu(\bi{u},\bi{v}|\bi{a},\bi{b}) \bi{v}=0$ in order to reproduce the correct quantum mechanical probability, and is 
otherwise arbitrary. The model violates both Uncorrelated Choice and Reducibility of Correlations, and 
satisfies Setting-Independence and Compliance with Malus's Law. 
\subsection{Hall model.}
Hall\cite{Hall2010} proposed the following deterministic model
\begin{eqnarray}
\label{eq:Hallmu}
\mu(\bi{u},\bi{v}|\bi{a},\bi{b})=
\delta(\bi{u}+\bi{v}) 
\frac{1-f(\bi{u},\bi{v}, \bi{a},\bi{b})}
{8\arccos{f(\bi{u},\bi{v}, \bi{a},\bi{b})}},\\
\label{eq:Hallp}
P_{\sigma,\tau}(\bi{u},\bi{v}, \bi{a},\bi{b})=
\frac{1}{4}\left[1+\sigma\mathrm{sgn}(\bi{u}\cdot\bi{a})\right] 
\left[1+\tau\mathrm{sgn}(\bi{v}\cdot\bi{b})\right],
\end{eqnarray}
where 
\begin{equation}
f(\bi{u},\bi{v}, \bi{a},\bi{b})=
\mathrm{sgn}(\bi{u}\cdot\bi{a})
\mathrm{sgn}(\bi{v}\cdot\bi{b})\ \bi{a}\cdot\bi{b} . 
\end{equation}
Hall's model clearly violates only the Uncorrelated Choice hypothesis, and it thus so 
using the minimum amount of correlation\cite{Hall2010}. The model does not 
comply with Malus's law since $P_{\sigma}(\bi{u},\bi{v}, \bi{a},\bi{b})=[1+\sigma\mathrm{sgn}(\bi{u}\cdot\bi{a})]/2$. 
\subsection{Di Lorenzo model.}
We proposed \cite{DiLorenzo2012b} the following model: 
\begin{eqnarray}
\label{eq:DiLorenzomu}
\mu(\bi{u},\bi{v}|\bi{a},\bi{b})=
\frac{1}{4} \sum_{\bi{n}=\pm \bi{a},\pm \bi{b}}
 \delta(\bi{u}-\bi{n}) \delta(\bi{v}+\bi{n})\\
\label{eq:DiLorenzop}
P_{\sigma,\tau}(\bi{u},\bi{v},\bi{a},\bi{b})=
\frac{1}{4}\left(1+\sigma \bi{u}\cdot\bi{a}\right)
\left(1+\tau \bi{v}\cdot\bi{b}\right) \ .
\end{eqnarray}
The model can be generalized by breaking the symmetry in Eq.~\eref{eq:DiLorenzomu}, 
\begin{equation}
\label{eq:DiLorenzomu2}
\mu(\bi{u},\bi{v}|\bi{a},\bi{b})=
\sum_{\bi{n}=\pm \bi{a},\pm \bi{b}}
 w_\bi{n}\delta(\bi{u}-\bi{n}) \delta(\bi{v}+\bi{n}),
\end{equation}
with $\sum_{\bi{n}}w_\bi{n}=1$ and $w$ non-negative weights.
This model is the only one\footnote{A former attempt was made in Ref.~\cite{DeZela2008}, where however 
the distribution $\mu$ was not normalized \cite{DiLorenzo2012e}.} that satisfies at the same time 
Setting-Independence, Malus's law, and Reducibility of Correlations, while it violates only 
Uncorrelated Choice. 
\bibliographystyle{iopart-num}
\section*{Bibliography}
\providecommand{\newblock}{}

\end{document}